\documentclass[a4paper, 12pt]{article}
\usepackage[english]{babel}
\usepackage{graphicx}
\usepackage{amssymb}

\begin{document}
\title{\bf{The frequency distribution of the height above the
Galactic plane for the novae}}

\author{\it{M.A.Burlak}}

\date{Sternberg Astronomical Institute, Moscow\\}

\renewcommand{\abstractname}{ }

\maketitle

\begin{abstract}

In order to examine the hypothesis of the existence of two
different kinds of nova populations in the Galaxy-- 'disk' novae
and 'bulge' novae -- the frequency distribution in the z-direction
was obtained for 64 novae. The fact that large number of fast
novae related to disk novae are found at a significant distance
from the Galactic plane (up to $z\!\sim\!3700$~pc) can't result
from photometric measurements errors. Slow novae considered to
belong to bulge novae show more close concentration to the
Galactic plane ($z\!\leqslant\!1700$~pc). A Kolmogorov-Smirnov
test run on the data showed that the two populations hypothesis
probability amounts to 95.56\%.



{\it Send offprint requests to} M.Burlak, e-mail:
mburlak@sai.msu.ru
\end{abstract}

\newpage

\subsection*{Hypothesis of the existence of two different kinds of
nova populations: disk and bulge novae.}

Traditionally the study of novae has been based on the assumption
that different kinds of stellar populations have different spatial
distributions in the galaxies. This idea has been used to find out
what kind of population novae progenitors belong to. The nova
population appeared to possess discordant features: by the
concentration towards the Galactic center it is similar to
Population~II objects, but by the concentration towards the
Galactic plane it resembles Population~I objects. As a result
novae have been considered either disk objects or bulge/thick disk
objects alternately.

Last years almost every study of the spatial distribution of novae
in the Galaxy and in neighbouring star systems have been carried
out in terms of the concept proposed by D\"{u}rbeck (1990) that
assumes the existence of two physically different kinds of nova
populations with differing galactic distributions and whose
progenitors probably differ from each other in a qualitative
sense. Bright and fast novae refer to disk population, slow and
faint novae belong to bulge (or thick disk). The differences
between the classes of novae account for the different nature of
the nova progenitors. Bright and fast novae are believed to be
associated with relatively massive white dwarfs
($M_{WD}\geqslant1M_{\odot}$) whereas faint and slow novae have
less massive progenitors ($M_{WD}\leqslant1M_{\odot}$). From the
physical point of view the matter is that the more massive the
white dwarf, the smaller the mass of the accreted envelope
required to produce a TNR, the more violent the outburst (i.e.
higher luminosity and expansion velocities, shorter $t_{3}$ -- the
time required to decline by 3 magnitudes from maximum), the larger
the mass fraction ejected as a discrete shell, the smaller the
mass ejected in the form of subsequent wind.

Della Valle et al.(1992) showed that the nova rate of decline
related to the mass of an underlying white dwarf correlates with
the nova location in the Galaxy. Using rates of decline and
outburst amplitudes for 93 galactic novae Della Valle et al.(1992)
found that fast (and bright) novae tend to concentrate towards the
direction of galactic anti-center, and slow (and faint) novae are
more frequently observed in the direction of the galactic center:
disk novae contribute more significantly when counting novae in
the direction of the galactic anti-center and bulge novae are more
numerous in the opposite direction. Besides, 19 novae with well
established distances were used to determine the frequency
distribution of the height above the galactic plane which showed
that fast novae more closely concentrating to the galactic plane
are found mainly at the heights $|\,z|\!\leqslant\!100$~pc.
Whereas the concentration of slow novae is not so close and they
extend up to $z\!>\!1000$~pc. From this the classification of
novae in two different classes, 'disk' novae and 'bulge' novae,
follows. Sometimes the latter are referred to by 'thick disk'
novae obviously taking into consideration the fact that bulge is
implied to be a more compact system that not expands to the radius
where novae are found. However in spite of some discrepancy these
objects are more often called bulge novae to emphasize their
concentration to the galactic center.

{\sloppy Later Della Valle \& Livio (1998) found that there are
spectroscopic differences between disk and bulge novae and that
the classification of novae based on the galactic distribution
nearly coincides ($\sim80$\%) with the spectral classification
introduced by Williams (1992), i.e. disk novae usually belong to
He/N class, whereas bulge novae are among Fe~II class members.

}

The two-nova-populations concept is widely used while the spatial
distribution of novae in other galaxies is analyzed. It predicts
that in bulge-dominated galaxies the percentage of faint slow
novae has to be larger whereas in disk-dominated galaxies bright
fast novae are more likely to erupt. So, Shafter \& Irby (2001)
consider the greater part of novae in M31 to belong to bulge
population (up to 70\% and not less than 50\%). For the Galaxy the
upper limit for the percentage of disk novae is thought to be 30\%
(Della Valle \& D\"{u}rbeck, 1993).

It's worth mentioning that the conclusion that classical novae
divide into two different classes has been drawn by various
authors on considering a little number of well-studied classical
novae whose light curves, distances, extinction, spectral
evolution are known. Such stars are not numerous. As a rule these
ones are close objects for which it became possible to measure the
angular expansion of the envelope. Having examined a sample of 27
novae Della Valle \& Livio (1998) pointed out the existence of
systematic spectroscopic differences between disk novae and bulge
novae. 10 of them were classified as He/N novae or hybrid objects
with mixed characteristics, and the residuary 17 were ascribed to
Fe~II class. Della Valle \& Livio (1998) applied a
Kolmogorov-Smirnov test of the uniformity of two samples to verify
the existence of two populations of novae with different
z-distributions. Significance level of the assumption appeared to
be $\gtrsim\!95\%$. Interestingly enough, in earlier paper
presented by the same research group (Della Valle et al, 1992)
where the z-distribution for only 19 novae was investigated the
difference in the distributions was significant at a level higher
than 99\%.

Single researchers do not confirm the existence of significant
difference between two groups of novae with various rates of
decline. The overwhelming belonging of novae in M31 to bulge was
questioned (Hatano et al, 1997). Sharov (1993) for the sample of
117 novae in M31 demonstrated that it is early to consider the
given hypothesis to be proven and that further study is required.

\subsection*{Construction of z-distribution for novae}

{\sloppy Present paper is the final part of the research of the
galactic novae exploded in 1986-2006. The first and the second
stages were described in papers published earlier: Burlak \&
Henden (2008) where the nova photometric parameters determination
with the use of visual light curves is discussed and Burlak (2008)
which concerns the calculation of distance and interstellar light
extinction for the given objects. The aim of the present study is
to analyze the frequency distribution of the height above the
Galactic plane for the novae. The calculations carried out in the
papers mentioned above yielded distance estimates for 64 galactic
novae. The height above the galactic plane was obtained using
distance and galactic latitude. In Table~\ref{ztab} the absolute
values of z are demonstrated. The $t_{3}$ estimates taken from
Burlak \& Henden (2008) are present here too. In order to examine
the question of whether the hypothesis of two nova populations
holds the stars were divided into two groups according to their
rate of decline: the group of fast novae consisted of 17 objects
with $t_{3}\!\leq\!20$~days, the rest 47 objects constituted the
group of slow novae. Given classification doesn't coincide with
the classification of Payne-Gaposchkin (1957), who places objects
with $t_{2}\!\leq\!25$~days ($t_{2}$ is the time it takes a nova
to decline by 2 magnitudes from maximum) among fast novae, but
fits the contemporary understanding of the nova phenomenon better
and therefore makes the comparison with other authors' results
more illustrative.

The histogram in Figure~\ref{zfig1} presents the frequency
distribution for the novae of two samples relative to the height
above the galactic plane. Filled area refers to the fast novae,
empty area corresponds to the slow novae. For comparison the
z-distribution for He/N and Fe~II novae obtained by Della Valle \&
Livio (1998) is presented in Figure~\ref{zfig2}. Though Della
Valle \& Livio (1998) plot this histogram to prove the
spectroscopic difference between disk and bulge novae the authors
conclude that speed classes actually correspond to the
spectroscopic classes.

}

\subsection*{Discussion}

\begin{figure}[!]
  \begin{center}
  \resizebox{13cm}{!}{\includegraphics{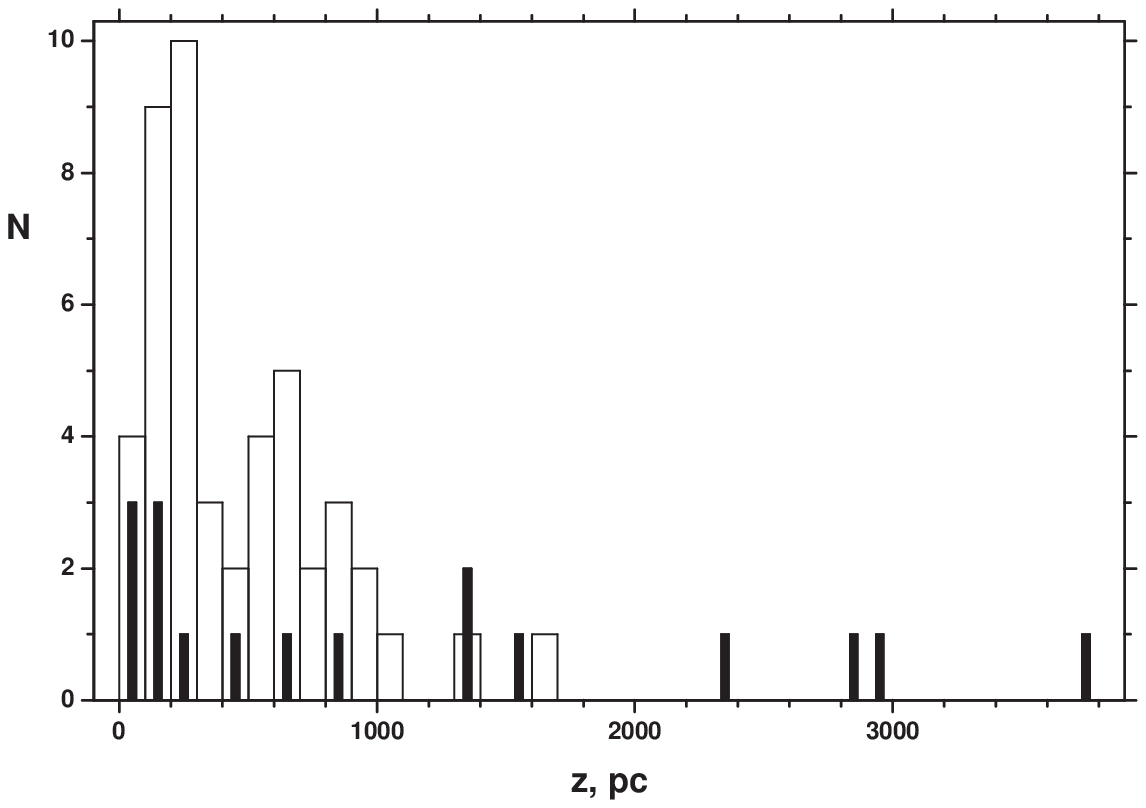}}
  \vspace{-0.8cm}
  \caption{Frequency distribution of the height above the Galactic plane for
  slow novae (unfilled columns) and fast novae (filled columns).}\label{zfig1}
  \resizebox{13cm}{!}{\includegraphics{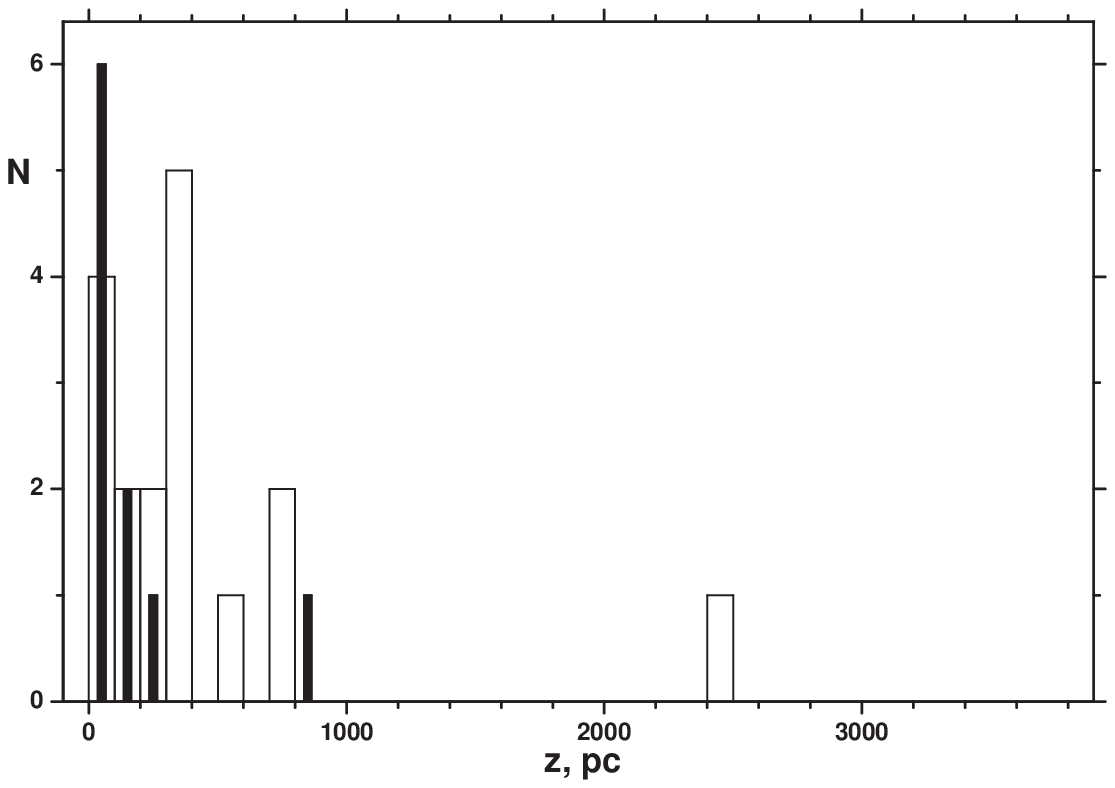}}
  \vspace{-0.8cm}
  \caption{Frequency distribution of height above the Galactic plane for Fe~II novae (unfilled
  columns) and He/N novae (filled columns) taken from Della Valle \& Livio (1998).}\label{zfig2}
  \end{center}
\end{figure}

One cannot say that the histogram in Figure~\ref{zfig1} agrees
well with the two-nova-population hypothesis. Fast novae do not
concentrate closely to the Galactic plane, only 6 of 17 fast novae
are found inside the $0\!\leq\!z\!\leq\!200$~pc strip, while the
rest fast novae are distributed homogeneously enough up to
$z\!\sim\!4000$~pc. And slow novae concentrate more close to the
Galactic plane and are not found at the heights $z\!>\!1700$~pc.
Obtained Z-distribution for slow novae corresponds with the
predictions of the discussed hypothesis to a greater extent.

Lets try to find out how the choice of interstellar light
extinction made in Burlak (2008) affected the form of histogram.
For novae located at $z\!>\!200$~pc light extinction and distance
values obtained with the maps of Schlegel et al. (1998) were
adopted. These maps yield the maximum light extinction in the line
of sight. Thus, if nevertheless interstellar extinction for a
given nova appears to be smaller than the maximum value then
distance for it will be greater than the adopted value.
Correspondingly the height above the Galactic plane will turn out
to be larger and the star will move to the right on the histogram
of z-distribution. So, the possible overestimation of the
extinction does not reduce the concentration to the Galactic plane
for the fast novae. The question arises then if the extinction
obtained with the maps of Sharov (1963) that take into account the
increase of interstellar extinction with distance appeared to be
underestimated for fast novae. Having compared the sets of
distances obtained according to the maps of Sharov (1963) and the
maps of Schlegel et al. (1998) one can see that there are only two
fast novae for which such a possibility exists, namely V~1494~Aql
and V~1187~Sco. But even though to adopt larger distances for both
novae they will only move from the range $100\!<\!z\!<\!200$~pc to
the range $0\!<\!z\!<\!100$~pc. The general form of the
distribution for the fast novae will not change. As follows from
the above the form of z-distribution for the fast novae does not
change in a qualitative sense when one set of distances is
replaced by the other in spite of significant differences between
two sets of distance values obtained with the aid of the
interstellar extinction maps derived by Sharov (1963) and by
Schlegel et al. (1998).

On the other hand incorrect distance modulus may lead to incorrect
distance value. Burlak \& Henden (2008) used the empirical
relation obtained by Cohen (1985)
$$
\textrm{M}_{V}=-10.66(\pm0.33)+2.31(\pm0.26)\times\lg t_{2},
$$
to get the absolute maximum magnitude. According to this formula a
20\% error in $t_{2}$ yields absolute magnitude error of
$0.18^{m}$ that changes distance by only 10\%. If there is a
detailed light curve for a given fast nova it is unlikely to make
a mistake larger than 1~day while estimating $t_{2}$ and such a
mistake will slightly affect the form of z-distribution for fast
novae. While estimating $t_{2}$ for slow novae some uncertainties
arise because of the light curve features. In exceptional cases
the discrepancy in the estimates of different researchers may rise
to hundreds of days. For example for V~723~Cas the following rate
of decline values were derived: $t_{3}=173$~days (Chochol \&
Pribulla, 1997), $t_{3}=778$~days (Evans et al., 2003),
$t_{3}=230$~days (Iijima, 2006), $t_{3}=779$~days (Burlak \&
Henden, 2008). If there are no flashes and deep light drops in the
light curve then it is possible to estimate $t_{2}$ and/or $t_{3}$
with adequate accuracy. One could say that errors in deriving rate
of decline do not influence the general form of z-distribution for
slow novae.

Underestimating the observed maximum brightness has the most
considerable effect on the distance value and therefore on the
z-distance value. The missing of the maximum brightness is
particularly unwanted in the case of fast novae but exactly fast
novae suffer from this more often. In Figure~~\ref{zfig1}\quad 7
fast novae are found at the height $z\!>\!1000$~pc, namely
V~838~Her, V~2295~Oph, DD~Cir, V~4332~Sgr, V~4160~Sgr, V~4739~Sgr,
V~2487~Oph. Their light curves generated from visual observations
of AAVSO members (Burlak \& Henden, 2008) do not rule out that the
maximum brightness may have been lost. For V~838~Her we managed to
find a detailed enough V light curve (Woodward et al., 1992) that
yields $5.3^{m}$ for the observed magnitude of maximum luminosity
for the nova. From the AAVSO light curve the observed maximum
magnitude for V~838~Her is equal to $7.6^{m}$. In the case of
V~838~Her the AAVSO observers proceeded to monitoring the star one
or two days past maximum but as V~838~Her was one of the fastest
novae its brightness declined by more than $2^{m}$ in this period.
If to adopt the maximum visual brightness for V~838~Her equal to
$5.3^{m}$ then the distance and hence z-distance reduce by a
factor of 3. Underestimating the nova maximum brightness by even
$1^{m}$ increases the distance by a factor of 1.5. So the omission
of maximum light influence strongly the form of z-distribution for
fast novae. In the case of slow novae a delay of 1-2 days or even
more will hardly affect the maximum light estimate and therefore
will not have an impact on the form of spatial distribution.

The omission of maximum not only causes underestimate of the
maximum brightness. As fast novae decline not uniformly just after
maximum but reduce the rate of decline then if to miss light
maximum then the time $t_{2}$ turns out to be overestimated.
Woodward et al. (1992) measured the time for a two-magnitude
fading and found $t_{2}\!\sim\!2$~days for V~838~Her whereas the
AAVSO light curve yields $t_{2}=5.5$~days. Woodward et al. (1992)
used the same empirical maximum magnitude - rate of decline
relation as Burlak \& Henden (2008) and obtained $-9.8^{m}$ for
the absolute magnitude of maximum luminosity (Burlak \& Henden
(2008) obtained $-8.9^{m}$). Thus if to adopt that AAVSO observers
missed the maximum light then Burlak \& Henden (2008) have
overestimated the observed distance modulus for V~838~Her by
$1.4^{m}$ since maximum brightness underestimate and $t_{2}$
overestimate have partially compensated each other. As a result
distance for V~838~Her and z-distance as well turn out
overestimated by a factor of 1.5 that affects substantially the
position of the nova in the histogram in Figure~\ref{zfig1}. If
distance for V~838~Her is reduced by 1.5 times then z-distance
turns out to be 850~pc but nevertheless the star does not fall
into a thin layer close to the Galactic plane. For the rest 6
stars located higher than 1000~pc above the Galactic plane it is
not possible to estimate the accuracy of photometric parameters
determination. But all of them being very fast novae are subjected
to both of the effects mentioned above. Fortunately one of them
increases observed distance modulus and another reduces it, so
there is no reason to suppose the distance error caused by
observational incompleteness to be large. Assuming distance
modulus to be overestimated by $2^{m}$ for all fast novae will
reduce z-distances for them only by a factor of 2.5 but
nevertheless the stars will remain in the range
$500\!<\!z\!<\!1500$~pc.

It's necessary to point out one more effect caused by the omission
of maximum light and the overestimate of $t_{2}$. For a fast nova
with true $t_{2}\!\leq\!20$~days overestimate of $t_{2}$ will
change its speed class from the fast one to the slow. There is a
large number of slow novae in the range $0\!<\!z\!<\!300$~pc. If
some of them prove to be fast novae in fact then the form of
z-distribution for fast novae may change noticeably in the
vicinity of the Galactic plane. But this can't reduce the number
of fast novae at large heights. On the other hand if some of the
slow novae at small z-distance turn out fast novae the general
form of z-distribution for slow novae will not change.

So, observational uncertainties have little impact on
z-distribution for slow novae whereas the characteristics of
z-distribution for fast novae depend strongly on photometric
parameters estimation accuracy. But photometric measurements
errors fail to explain the fact that there is a large number of
fast novae far from the Galactic plane.

It's interesting to note that seven fast novae mentioned above
have Galactic latitude larger than $6^{\circ}$. All other fast
novae (except very bright V~382~Vel) are located lower than
$5^{\circ}$ and experience considerable interstellar light
extinction. A great amount of fast novae at low Galactic latitudes
must remain undetected because of high star density and high
column density of interstellar medium close to the Galactic plane.
In other words the matter is not that the obtained z-distribution
contains too many fast novae at a large height above the Galactic
plane but that there are too little of them close to it. This fact
in its turn accounts for strongest observational selection.

{\sloppy To make up final conclusion if the obtained results agree
with the two-nova-populations hypothesis a Kolmogorov-Smirnov test
was performed (Hollander \& Wolfe, 1983). It showed the difference
in z-distributions to be significant at a level of 95.56\% that
coincides with the value obtained by Della Valle \& Livio (1998)
for a sample of 27 novae. But the test just points out
non-identity of two samples and says nothing about the
distribution laws features and the observational data is not
enough to find them out.

}

\subsection*{Conclusions}

\begin{enumerate}
{\sloppy
    \item Z-distribution for Galactic novae exploded after
    1986 did not show significant concentration of fast novae to the
    Galactic plane. On the contrary large part of them (7 of 17)
    was found higher than 1000~pc above the plane.
    \item The z-distribution form for fast novae depends strongly
    on photometric parameters estimation accuracy. But a large
    amount of fast novae at a considerable distance from the
    Galactic plane can't be accounted for inaccuracy of photometric
    measurements alone.
    \item Relatively small number of fast novae located close to
    the Galactic plane arises from strong observational selection
    effect caused by interstellar light extinction.
    \item A Kolmogorov-Smirnov test supports the hypothesis of two
    novae populations with different z-distributions at a
    significant level of 95.56\%.

}
\end{enumerate}

\newpage

REFERENCES

\bigskip
{\sloppy
\begin{enumerate}

    \item Burlak, M.A., Henden, A.A. 2008, PAZh, {\it in press}
    \item Burlak, M.A. 2008, PAZh, {\it in press}
    \item Chochol, D., Pribulla, T. 1997, Contrib.
    Astron. Obs. Skalnat\`{e} Pleso, 27, 53
    \item Cohen, J.G. 1985, ApJ, 292, 90
    \item Della Valle, M., Bianchini, A., Livio, M., Orio, M. 1992, A\&A, 266, 232
    \item Della Valle, M., D\"{u}rbeck, H.W. 1993, A\&A, 271, 175
    \item Della Valle, M., Livio, M. 1998, ApJ, 506, 818
    \item D\"{u}rbeck, H.W. 1990, in {\it 'The Physics of Classical
    Novae'}, eds. A. Cassatella and R. Viotti (Springer-Verlag Berlin Heidelberg New York)
    \item Evans, A., Gehrz, R.D., Geballe, T.R. 2003, AJ, 126, 1981
    \item Hatano, K., Branch, D., Fisher, A., Starrfield, S. 1997,
    ApJ, 487, L45
    \item Hollander, M., Wolfe, D.A. 1983, in {\it 'Nonparametric Statistical
    Methods'} (Moscow: Finances \& Statistics)
    \item Iijima, T. 2006, A\&A, 451, 563
    \item Payne-Gaposchkin, C. 1957, in {\it 'The Galactic
    Novae'}, Amsterdam: North-Holland Co
    \item Shafter, A.W., Irby, B.K. 2001, ApJ, 563, 749
    \item Sharov, A.S. 1963, AZh, 5, 900
    \item Sharov, A.S. 1983, PAZh, 19, 18
    \item Schlegel, D.J., Finkbeiner, D.P., Davis, M. 1998, ApJ, 500, 525
    \item Williams, R.E. 1992, AJ, 104, 725
    \item Woodward, C.E., Gehrz, R.D., Jones, T.J., Lawrence, G.F. 1992, ApJ, 384, L41
\end{enumerate}
}

\newpage
\begin{table}[!t]
 \begin{center}
 \caption{Height above the Galactic plane}\label{ztab}
 \begin{tabular}{|l r r||l r r|}
   \noalign{\smallskip} \hline \noalign{\smallskip}%
    Nova & $t_{3}$, days & $z$, pc & Nova & $t_{3}$, days & $z$, pc \\
   \noalign{\smallskip} \hline \noalign{\smallskip}
    V 4739 Sgr & 3.5 & 2900 & V 1419 Aql & 33 & 290\\
    V 2361 Cyg & $\geqslant\!7$ & 610 & V 1663 Aql & 34 & 31\\
    V 838 Her & 8.7 & 1300 & V 2264 Oph & 38 & 670\\
    V 2275 Cyg & 10 & 73 & V 4157 Sgr & 41 & 290\\
    V 4160 Sgr & 10 & 2800 & V 4169 Sgr & 44 & 850\\
    V 2487 Oph & 10 & 3700 & V 5116 Sgr & 45 & 1300\\
    V 4332 Sgr & 10 & 2300 & V 1141 Sco & 46 & 490\\
    V 5115 Sgr & 14 & 820 & V 1186 Sco & 46 & 640\\
    V 4643 Sgr & 14 & 18 & QV Vul & 47 & 270\\
    V 4444 Sgr & 14 & 490 & V 1974 Cyg & 48 & 200\\
    V 382 Vel & 15 & 110 & V 868 Cen & 48 & 130\\
    DD Cir & 15 & 1500 & V 842 Cen & 49 & 51\\
    V 1494 Aql & 16 & 170 & V 2274 Cyg & 50 & 170\\
    V 2295 Oph & 16.3 & 1300 & V 2574 Oph & $\geqslant\!50$ & 640\\
    V 1187 Sco & 18 & 180 & V 888 Cen & 54 & 240\\
    CP Cru & 18 & 89 & V 475 Sct & 55 & 320\\
    V 4171 Sgr & 20 & 280 & V 4633 Sgr & 62 & 760\\
    V 4743 Sgr & 22 & 1050 & V 827 Her & 64 & 840\\
    V 4742 Sgr & 23 & 170 & V 1039 Cen & 71 & 160\\
    V 2362 Cyg & 23 & 280 & V 2214 Oph & 80 & 670\\
    V 444 Sct & 23 & 950 & V 574 Pup & 85 & 170\\
    V 463 Sct & \textbf{>}23 & 1600 & V 4361 Sgr & 86 & 120\\
    V 5114 Sgr & 24 & 930 & V 705 Cas & 96 & 130\\
    V 1142 Sco & $\geqslant\!25$ & 270 & V 2573 Oph & 99 & 500\\
    V 1493 Aql & 26 & 170 & BY Cir & 101 & 150\\
    V 4741 Sgr & 28 & 530 & V 1425 Aql & 119 & 220\\
    V 1188 Sco & 28 & 220 & V 4642 Sgr & 125 & 240\\
    V 2576 Oph & 29 & 780 & V 443 Sct & 135 & 220\\
    V 4327 Sgr & 30 & 830 & V 351 Pup & 284 & 47\\
    V 4740 Sgr & 30 & 470 & V 2540 Oph & 305 & 570\\
    LZ Mus & 32 & 510 & V 445 Pup & 320 & 92\\
    V 2313 Oph & 32 & 690 & V 723 Cas & 779 & 390\\

   \noalign{\smallskip} \hline \noalign{\smallskip}
 \end{tabular}
 \end{center}
\end{table}

\newpage

\end{document}